\documentclass[%
reprint,
 amsmath,amssymb,
]{revtex4-1}

\usepackage{bm} 
\usepackage{graphicx}
\usepackage{amsfonts}
\usepackage{amscd}
\usepackage{enumerate}
\usepackage{subfigure}
\usepackage{amsthm}
\usepackage{dcolumn}
\usepackage{amsmath,amssymb}

\usepackage{txfonts}

\usepackage[breaklinks]{hyperref}

\begin{document}
\title{Multi-channel parallel continuous variable quantum key distribution with Gaussian modulation}
\author{Jian Fang}
\email{aceofspades@sjtu.edu.cn}
\author{Peng Huang}
\email{huang.peng@sjtu.edu.cn}
\author{ Guihua Zeng}
\email{ghzeng@sjtu.edu.cn}
\affiliation{State Key Laboratory of Advanced Optical Communication Systems and Networks,
Shanghai Jiaotong University, Shanghai 200240, China}

\begin{abstract}

We propose a novel scheme for continuous variable quantum key distribution(CV-QKD) using the subcarrier multiplexing technique which was employed in microwave photonics. This scheme allows to distribute $N$ channels independent Gaussian modulated CV-QKD in parallel with one laser source and several phase modulators. We analyze the influence of nonlinear signal mixing and the security in the asymptotic limit. Results indicate that by using this multiplexing technique, each channel will be introduced a non-Gaussian extra source noise, resulting slightly short of the maximum transmission distance, while the total secret key rate can be considerably increased. This scheme could also be used for key distribution for multi-users and combined with the WDM devices it has the potential to construct a CV-QKD network.

\end{abstract}

\maketitle

\section{introduction}

Quantum key distribution (QKD), as a major practical application of quantum information, allows two distant parties to share a common secret key for cryptography in an untrusted environment\cite{Gisin2002,Scarani2009,Weedbrook2012-1}. Its security is guaranteed by the laws of quantum mechanics. QKD with continuous-variable, which is an alternative to the single-photon-based discrete-variable quantum key distribution(DV-QKD), encodes information on the quadratures of a Gaussian state\cite{Grosshans2002}. Any eavesdropping will introduce extra noise between two legal communication parties, who can realize Eve's existence by detecting the excess noise.

The Gaussian modulated CV-QKD protocols, which have been experimentally demonstrated both in laboratory\cite{Nature,Lodewyck2007,Bing2007,Jouguet2013} and field test\cite{Fossier2009}, are proven secure against collective attacks\cite{Raul2006,Navascues2006} and coherent attacks\cite{Renner2009,Furrer2012}. They use homodyne detectors instead of the single-photon counters employed in DV-QKD systems, and are more attractive from a practical point of view. By using the multi-dimensional reverse reconciliation\cite{Leverrier2008,Jouguet2011}, the secure transmission distance can be extended to as long as 80km\cite{Jouguet2013}. 

Most of the existing CV-QKD systems are pulsed systems and the secret key bit rate $R$ can be expressed as $R=f_{\text{rep}}K$ where $f_{\text{rep}}$ denotes the pulse repetition rate and $K$ is the secret key  rate(bit per pulse). Comparing with classical communication systems, the secret key bit rate of CV-QKD is still low, ranging from several bits/s to hundreds of kbits/s at the distance of more than 25km\cite{Lodewyck2007,Bing2007,Fossier2009,Jouguet2013}. There are three possible approaches to solve this problem. The first one is to improve the secret key rate $K$, which is a diminishing  function of the transmission distance and sensitive to the excess noise. So at long transmission distance the secret key rate is much smaller than $10^{-3}$ bit per pulse. The second approach is to increase the frequency of pulse repetition rate, which requires faster data acquisition cards, wider bandwidth of quantum detectors and the higher speed of post-processing procedure. 

The third approach is to use the multiplexing technique, which allows to deliver $N$ independent secret keys in a single fiber. This method was initially realized by using the wavelength division multiplexing(WDM) devices, which needs $N$ coherent laser sources with different central frequencies, and each channel needs individual amplitude and phase modulators. So the WDM scheme can be just deemed as a combination of several independent QKD systems. Interestingly, the subcarrier multiplexing technique, which was employed in the field of microwave photonics and the radio-on-fiber(ROF) systems, has been proven useful in the BB84 protocol recently\cite{Capmany2006,Capmany2009,Capmany2012}. In their scheme, the sender Alice randomly encodes one of the discrete phases $\{0,\pi/2,\pi,3\pi/2\}$ on several radio-frequency(RF) oscillators and the receiver Bob decodes them by using the oscillators of the identical frequencies as Alice. This method requires a frequency locking module between two distant parties, which may increase the complexity of the whole system.

Inspired by this method, we propose a new scheme to employ the subcarrier multiplexing technique in the Gaussian modulated CV-QKD protocol. Different from Ref.\cite{Capmany2012}, each RF oscillator is modulated with continuous phase and amplitude information, while the system does not require frequency locking. In our proposal, the subcarrier frequencies are evenly separated. We consider the influence of nonlinear signal mixing and analyze the security against collective attacks. The results are attractive. The Gaussian modulation of each channel will introduce a non-Gaussian extra noise which is generated from the continuous information modulated on other channels. This extra noise is proportional to Alice's modulation variance and cannot be neglected. 

We evaluate the secret key bit rate in the asymptotic limit for both each channel and the whole system. Results indicate that the extra source noise will slightly reduce of the maximum transmission distance, while the total secret key rate can be considerably increased. Also each channel could generate the independent secret keys at the same time, meaning the key distribution in the multi-channel system is parallel. This scheme could be used to distribute keys to multi-users. Combined with current WDM technique, the multi-channel scheme has the potential to construct a CV-QKD network.

This paper is structured as follows. In Sec.\ref{S2}, a brief review of the Gaussian protocol is given. In Sec.\ref{S3}, we describes the principle of the multi-channel scheme. In Sec.\ref{S4} the effect the nonlinear signal mixing is studied to derive the expression of the extra source noise due to intermodulaton. We construct the entanglement-based scheme in Sec.\ref{S5} and use it to analyze the security under collective attacks. The results of numerical simulation and the discussion are provided in Sec.\ref{S6} and the conclusions are drawn in Sec.\ref{S7}.

\section{Brief review of the Gaussian protocol}\label{S2}

The Gaussian CV-QKD protocols are based on the Gaussian modulation of a Gaussian state of light which could be coherent state\cite{Grosshans2002}, squeezed state\cite{Cerf2001} or thermal state\cite{Weedbrook2012}. From practical point of view, the most suitable protocol for experimental demonstration is the Gaussian modulated coherent state(GMCS) quantum cryptography protocol, which was proposed by F. Grosshans and P. Grangier in 2002\cite{Grosshans2002}. The GMCS protocol requires two independent Gaussian distributed modulation with the identical variance $V_A$ on the $x$ and $p$ quadratures of a coherent state. Expressed in terms of phase and amplitude, the distribution corresponds to a uniform modulation of the phase in $[0,2\pi]$ and a Rayleigh distribution of amplitude with the probability density function $Ray(\sigma=\sqrt{V_A})$ where
\begin{equation}
z\sim Ray(\sigma) = \frac{z}{\sigma^2}e^{-\frac{z^2}{2\sigma^2}}.
\end{equation}

Alice's modulated states are sent to Bob through the quantum channel. The quantum channel is featured by the transmission efficiency $T$ and excess noise $\epsilon$, resulting in a noise variance of $1+T\epsilon$ at Bob's input. Different from DV-QKD protocol, the GMCS protocol requires a strong local oscillator(LO). In order to ensure the LO and signal light having the identical mode, both the LO and signal light are generated by Alice using an unbalanced beam splitter(BS) from the same pulsed coherent laser. Then Alice applies time multiplexing and polarization multiplexing techniques to let the LO and signal propagate in the same fiber.

When Bob receives the states, he first uses the demultiplexing devices to separate the LO and quantum signal. Then he performs homodyne detection, randomly measuring the $x$ or $p$ quadrature.  To measure the quadrature $p$, he dephases the local oscillator by $\pi/2$. Then Alice and Bob performs classical data processing, including the reconciliation and privacy amplification, and finally share the  common secret keys from the accumulated data.

\section{Description of the multi-channel scheme}\label{S3}

\begin{figure*}[htp1]
\includegraphics[width=16cm]{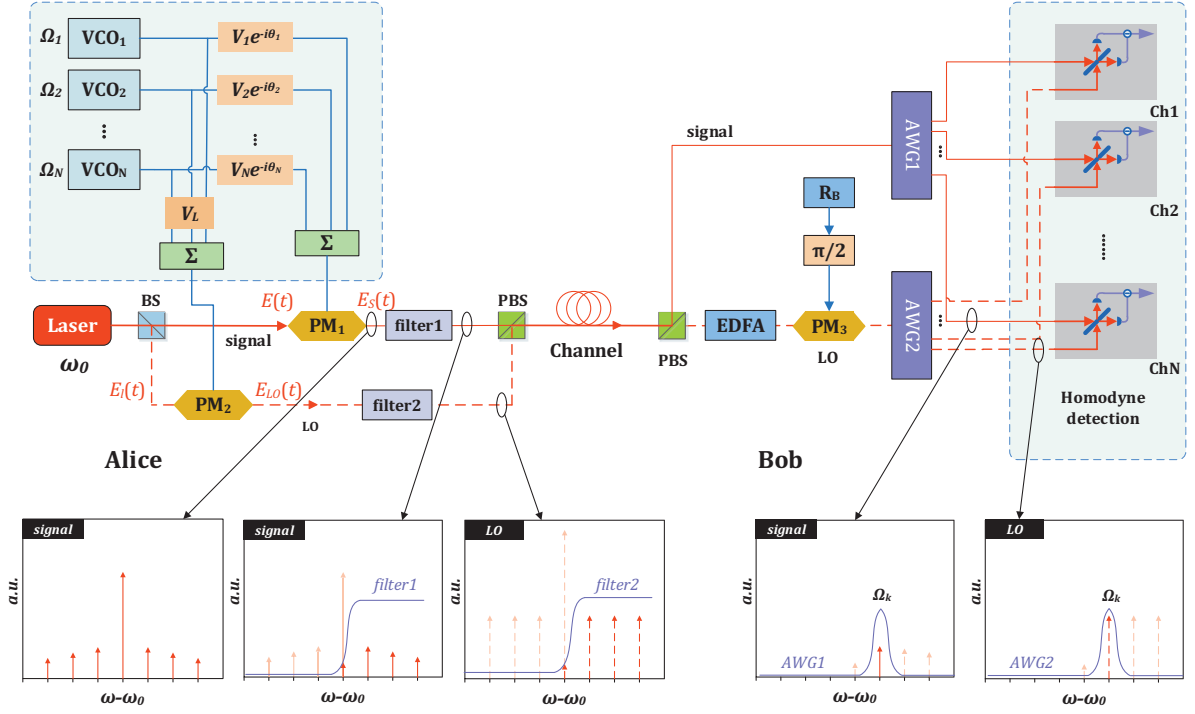}
\caption{\label{LAYOUT}(Color online) System layout of the multi-channel parallel CV-QKD scheme. The signal light is in red solid line and the local oscillator light is in red dashed line.}
\end{figure*}

The operation principle of the multi-channel scheme is shown in Fig.\ref{LAYOUT}. Alice uses a beam splitter(BS) to separate the coherent pulsed laser source centered at $\omega_0$ into two beams: one is weaker and the other is stronger. The weaker beam is prepared for generating quantum signals while the stronger beams is used for local oscillators(LO). The optical field is assumed to be quasi-monochrimatic, i.e. the  spectral width $\Delta\omega$ of the pulse spectrum is much smaller than its central frequency $\omega_0$. This assumption is always satisfied when the pulse width is  larger than $0.4$ps\cite{AgrawalBook}. The quantized single mode electric field $\hat{E}(t)$ of the signal light before modulation can be expressed as $\hat{E}(t) = \hat{E}^+(t)+\hat{E}^-(t)$ where
\begin{equation}\label{ET}
\hat{E}^+(t)= i \sqrt{\frac{\hbar \omega_0}{2\epsilon_0V_0}}\hat{a}(0)e^{-i\omega_0t}
\end{equation}
and $\hat{E}^-(t)=[\hat{E}^+(t)]^\dag $. $\epsilon_0$ is the dielectric permittivity and $V_0$ denotes  the mode volume. $i$ is the imaginary unit. $E^+(t)$ and $E^-(t)$ denote the positive-frequency and negative-frequency components of the quantized electric field, respectively. $\hat{a}(0)$ is the dimensionless complex amplitude operator which can decomposed by two quadratures  as $\hat{a}(0)=\hat{X}+i\hat{P}$. Since  $\hat{E}^+(t)$ and $\hat{E}^-(t)$ are conjugated, we only  need to consider the positive-frequency component in following discussions, while the identical results could be obtained from the negative one by similar methods. 

Signal light is externally modulated through a phase modulator(PM${}_1$) by $N$ radio-frequency(RF) subcarriers. Each subcarrier, which is generated from a voltage control oscillator(VCO) of frequency $\Omega_k,k\in\{1,2...,N\}$, is independently amplitude-modulated by a Rayleigh distributed random number $V_k$ and phase-modulated by a uniform distributed random number $\phi_k\in[0,2\pi]$. All the modulated RF signals  are combined together  with a bias voltage $V_b$. Then the voltage applied on PM${}_1$ is expressed as
\begin{equation}\label{VS}
V_S(t)=-\sum_{k=1}^NV_k\cos(\Omega_kt+\phi_k)-V_b.
\end{equation}
After Alice's phase modulation, the positive-frequency component turns to:
\begin{eqnarray}\label{EST}
\hat{E}_{S}^+(t)&=&\hat{E}^+(t)\exp\left[-\frac{i\pi V(t)}{V_{\pi}}\right]\nonumber\\
&=&\hat{E}^+(t) e^{i\theta_b}\exp{\left[i\sum_{k=1}^{N}m_k\cos(\Omega_kt+\phi_k)\right]},
\end{eqnarray}
where $m_k=V_k\pi /V_{\pi}$, $\theta_b=V_b/V_{\pi}$. $V_{\pi}$ is the half-wave voltage of the phase modulator.  This equation can be written as Taylor expansion with following form
\begin{eqnarray}\label{ES}
\hat{E}_S^+(t)
&=&\hat{E}^+(t)e^{i\theta_b}\sum_{n=0}^{+\infty}\frac{i^n}{n!}\left[\sum_{k=1}^{N}m_k\cos(\Omega_kt+\phi_k)\right]^n\nonumber\\
&\cong&\hat{E}^+(t)e^{i\theta_b}\Big[1+\frac{i}{2}\sum_{\substack{k=-N\\k\neq0}}^Nm_ke^{-i(\Omega_kt+\phi_k)}\nonumber\\
&&+\frac{i^2}{8}\sum_{\substack{r,s=-N\\r,s\neq0}}^Nm_rm_se^{-i[(\Omega_{r}+\Omega_{s})t+\phi_r+\phi_s]}\Big],
\end{eqnarray}
where we define $m_{-k}=m_{k}$, $\Omega_{-k}=-\Omega_{k}$ and $\phi_{-k}=-\phi_{k}$ for any integer $|k|\in\{1,...,N\}$.
Here we use the second-order Taylor expansion because when $ m_k$ is small enough, e.g. $m_k\leq0.02$, the effect of higher-order terms is negligible\cite{Capmany2009}.
Then the signal field at frequency $\omega_0+\Omega_k$  is:
\begin{equation}
\hat{E}^+_S|_{\omega_0+\Omega_k}(t)=\frac{1}{2}\hat{E}^+(t)e^{-i\Omega_kt+i\theta_b}C_k,
\end{equation}
where
\begin{equation} \label{Ck}
C_k = im_ke^{-i\phi_k}+\frac{i^2}{4}\sum_{\substack{r,s=-N\\r,s\neq0\\\Omega_r+\Omega_s=\Omega_k}}^Nm_rm_se^{-i(\phi_r+\phi_s)}.
\end{equation}

Similar as the form of Eq.(\ref{ET}), we can rewritten the expression of $\hat{E}_S^+|_{\omega_0+\Omega_k}$ as following:
\begin{equation}
\hat{E}^+_S|_{\omega_0+\Omega_k}(t)=\sqrt{\frac{\hbar( \omega_0+\Omega_k)}{2\epsilon_0V}}\hat{a}_k(0)e^{-i({\omega_0+\Omega_k})},
\end{equation}
where $\hat{a}_k(0)=X_{S(k)}+iP_{S(k)}$ is a new dimensionless complex amplitude operator of mode $\omega_0+\Omega_k$ and can be expressed as:
\begin{equation}
\hat{a}_k(0) =\frac{1}{2} \sqrt{\frac{\omega_0+\Omega_k}{\omega_0}}C_ke^{i\theta_b}\hat{a}(0)\cong\frac{1}{2}C_ke^{-i\theta_b}\hat{a}(0).
\end{equation}

If we adjust the bias phase $\theta_b$ to $-\arg(\hat{\alpha}(0))-\pi/2$, then the $X$ quadrature of  $\hat{a}_k(0)$ becomes
\begin{eqnarray}\label{Xsk}
X_{S(k)}=\frac{\alpha_0}{2}m_k\cos(-\phi_k)+\frac{\alpha_0}{8}\sum_{\substack{r,s=-N\\r,s\neq0\\\Omega_r+\Omega_s=\Omega_k}}^Nm_rm_s\sin(\phi_{r}+\phi_{s}),
\end{eqnarray}
where $\alpha_0$ denotes the norm of $\hat{a}(0)$. The expression of $P_{(k)}$ can be obtained by interchanging the $\sin(\cdot)$ and $\cos(\cdot)$ functions in Eq.(\ref{Xsk}).
The compound signal then passes through an optical filter, which blocks all the subcarriers below the frequency $\omega_{0}$. This is because the subcarrier phase modulation in PM${}_1$ encodes the same information ($m_k$ and $\phi_k$) on both the subcarriers centered at $\omega_0+\Omega_k$ and  $\omega_0-\Omega_k$, while we only use the positive frequency components for quantum key distribution. 

The local oscillator(LO) light can also be generated by the similar way as the signal light. The difference is that the voltage applied on the LO's modulator does not contain randomly modulated information. Therefore the modulation voltage of LO can be written as:
\begin{equation}
V_{LO}(t) = -V_L \sum_{k=1}^N\cos(\Omega_kt)-V'_b,
\end{equation}
where $V_L$ is the amplitude voltage of each RF component and $V'_b$ is the bias voltage.
Since LO is usually much larger(typically $10^6\sim10^8$ times) than the quantum signal, it can be deemed as a classical optical field $E_{l}(t)=E_0e^{-i\omega_0t}$.
So the LO light after modulator PM${}_2$ is:
\begin{eqnarray}\label{ELO}
E_{LO}(t) &\cong&E_l(t)e^{i\psi_b}\Big[1+\frac{im_{L}}{2}\sum_{\substack{k=-N\\k\neq0}}^Ne^{-i\Omega_kt}\nonumber\\
&&+\frac{i^2m_{L}^2}{8}\sum_{\substack{r,s=-N\\r,s\neq0}}^Ne^{-i(\Omega_r+\Omega_s)t}\Big],
\end{eqnarray}
where $m_L=V_L/V_{\pi}$ and $\psi_b=V'_b/V_{\pi}$. If bias phase $\psi_b$ is adjust to $-\pi/2$, then the local oscillator light at frequency $\omega_0+\Omega_k$ is
\begin{equation}
E_{LO}|_{\omega_0+\Omega_k}(t)=\frac{1}{2}E_0e^{-i(\omega_0+\Omega_k)t}D_k,
\end{equation}
in which the parameter $D_k$ is:
\begin{equation} \label{Dk}
D_k =m_{L}+\frac{i}{4}M_2(N,k)m_{L}^2,
\end{equation}
where $M_2(N,k)$ is the numbers of terms satisfied $\Omega_{r}+\Omega_{s}=\Omega_k,(|r|,|s|\in\{1,...,N\})$. The values of $M_2(N,k)$ can be calculated by enumeration method. We also derive the analytical expression of $M_2(N,k)$ as
\begin{equation}\label{M2}
M_2(N,k)= 2N-k-\frac{3}{2}-\frac{1}{2}(-1)^k.
\end{equation}
For a given total channel numbers $N$, $M_2(N,k)$ satisfies the following inequation:
\begin{equation}\label{bound}
N-2\leq M_2(N,k)\leq 2N-2.
\end{equation}
When $m_L\leq 0.01$ and $N\leq 40$, the impact introduced by nonlinear signal mixing on $|D_k|$ is less than $1.5\%$, which could be neglected. Therefore the LO field of each channel has the constant value $E_{LO}|_{\omega_0+\Omega_k}=\frac{1}{2}E_0m_Le^{-i(\omega_0+\Omega_k)t}$ as shown in Fig\ref{LAYOUT}. In order to reduce the effect of  scattering and nonlinearity in the fiber, the LO power should not be high enough. As the signal light, the LO is also filtered the  subcarriers below $\omega_0$. Then the compound signal light and  LO light are polarization multiplexed  into the quantum channel and transmitting to Bob.

After receiving the states sent by Alice, Bob firstly separates the compound signal and LO by using a polarization beam splitter(PBS). Then he uses an Erbium-doped optical fiber amplifier(EDFA) to amplify the LO to a considerable power and randomly phases the LO by $\{0,\pi/2\}$ to select the measurement basis. The random basis choosing procedure is determined by a binary random number generator which Bob holds its result as $R_B$. He then uses two arrayed-waveguide gratings(AWG) to filter out each subcarrier and performs balanced homodyne detection on each pairs of signal and LO(both are centered at frequency $\omega_0+\Omega_k$). Finally, Alice and Bob perform classical data processing, including the reverse reconciliation and the privacy amplification. After these stages, they share $N$ channel independent secret keys and the key distribution is completed.

In this paper, we consider three different frequency spacing plans: the low-plan($N=5$) with evenly spaced(e.g. $\Omega_k=k\Omega_1$) channels from 5 to 25GHz, the medium-plan($N=15$) with evenly spaced channels from 2 to 30GHz and the high-plan($N=40$) with evenly spaced channels from 1 to 40GHz. From practical point of view, 40GHz phase modulators are currently commercially available and the ultra-narrow band AWG devices with 1GHz channel spacing have been experimentally demonstrated\cite{Takada2002}. So it is reasonable for us considering these cases.

\section{Extra source noise due to intermodulation}\label{S4}

According to the Eq.(\ref{Xsk}), the expression of $X_{S(k)}$ can be written as two parts:
\begin{equation}
X_{S(k)}=X_{A(k)}+\Delta X_{(k)},
\end{equation}
where
\begin{eqnarray}\label{DeltaX}
X_{A(k)} &=& \frac{\alpha_0}{2}m_k\cos(-\phi_k)\nonumber\\
\Delta X_{(k)} &=& \frac{\alpha_0}{8}\sum_{\substack{r,s=-N\\r,s\neq0\\\Omega_r+\Omega_s=\Omega_k}}^Nm_rm_s\sin(\phi_{r}+\phi_{s}).\nonumber\\
\end{eqnarray}
Since $m_k$ follows a Rayleigh distribution $R(\sigma)$ and $\phi_k$ follows a uniform distribution in $[0,2\pi]$, $X_{A(k)}$ follows the Gaussian distribution $\mathcal{N}(0,\sigma\alpha_0/2)$, which is identical to the modulation in Gaussian CV-QKD protocol. The second part, $\Delta X_{(k)}$, can be seen as a modulation  noise. Notice that
\begin{equation}
m_rm_s\sin(\phi_r+\phi_s) = x_rp_s+p_rx_s,
\end{equation}
where $x_r=m_r\cos\phi_r$, $p_r = m_r\sin\phi_r$, $x_s=m_s\cos\phi_s$ and $p_s = m_s\sin\phi_s$. For different $r$ and $s$, the random variables $x_r$, $p_r$, $x_s$, $p_s$ are mutually independent and follow the same Gaussian distribution $\mathcal{N}(0,\sigma)$. Hence the probability density function of $x_r p_s$(also for $x_s p_r$) is given by\cite{WeissteinOnline}
\begin{eqnarray}
f(z)&=&\int_{-\infty}^{+\infty}\int_{-\infty}^{+\infty}\frac{e^{-\frac{x^2}{2\sigma^2}}}{\sigma\sqrt{2\pi}}\frac{e^{-\frac{y^2}{2\sigma^2}}}{\sigma\sqrt{2\pi}}\delta(xy-z)\text{d}x\text{d}y\nonumber\\
&=&\frac{1}{\pi \sigma^2}K_0\left(\frac{|z|}{\sigma^2}\right),
\end{eqnarray}
where $K_0(\cdot)$ is the zero order modified Bessel function of the second kind and $\delta(\cdot)$ is the delta function. The mean values of the products are $\langle x_rp_s\rangle=\langle x_sp_r\rangle=0$ while the variances are $\langle (x_rp_s)^2\rangle=\langle (x_sp_r)^2\rangle=\sigma^4$. So the variance of the term $m_rm_s\sin(\phi_r+\phi_s)$ is
\begin{equation}
\langle(m_rm_s\sin(\phi_r+\phi_s))^2\rangle = \begin{cases}
2\sigma^2 & \text{for $r\neq s$}\\
4\sigma^2 & \text{for $r= s$}\\
\end{cases}
\end{equation}

As shown in Sec.\ref{S2}, the total number of combinations satisfied $\Omega_s+\Omega_r=\Omega_k$  is $M_2(N,k)$. For an odd integer $k$, the combination $2\Omega_s=\Omega_k$ does not exist, so $\Omega_k$ can only be decomposed by the combinations of $\Omega_r+\Omega_s, r\neq s$. Since $\Omega_r$ and $\Omega_s$ are commutable, the term $m_rm_s\sin(\phi_r+\phi_s)$ exists twice. Therefore $\langle\Delta X_{(k)}\rangle=0$ and the variance of $\Delta X$ can be expressed as
\begin{equation}
\langle\Delta X_{(k)}^2\rangle=\frac{\alpha_0^2}{64}\left[\frac{M_2(N,k)}{2}\times 4\times 2\sigma^4\right]=\frac{\alpha_0^2}{16}M_2(N,k)\sigma^4.
\end{equation}
When $k$ is an even integer, there is one combination $2\Omega_s=\Omega_k$. So in this case  $\langle\Delta X_{S(k)}^2\rangle$ becomes:
\begin{equation}
\langle\Delta X_{(k)}^2\rangle=\frac{\alpha_0^2}{64}\left[\frac{M_2(N,k)-1}{2}\times 4\times 2\sigma^4+4\sigma^4\right]=\frac{\alpha_0^2}{16}M_2(N,k)\sigma^4,
\end{equation}
which is identical with the case of odd $k$. The variance of $\Delta P_{(k)}$ can be derived in the similar way. We assign the variance of $X_{A(k)}$ as $V_A=\alpha_0^2\sigma^2/4$. Since $m_k\sim Ray(\sigma)$, the mean value of $m_k$ should be $\overline{m}_k=\sigma\sqrt{\pi/2}$. Then the variance of  variance of $X_{(k)}$ can be expressed in terms of $\overline{m}_k$ and $V_A$ as
\begin{equation}\label{ESK}
\epsilon_{S(k)}\triangleq \langle\Delta X_{(k)}^2\rangle=\langle\Delta P_{(k)}^2\rangle=
\frac{1}{2\pi}M_2(N,k)\overline{m}_k^2V_A,
\end{equation}
and finally we get
\begin{equation}\label{varxk}
\langle X_{S(k)}^2\rangle =\langle P_{S(k)}^2\rangle=V_A+\epsilon_{S(k)},
\end{equation}
where $\epsilon_{S(k)}$ is the noise variance due to intermodulation defined in Eq.(\ref{ESK}). From a physical point of view, the generation of $\epsilon_{S(k)}$ is the result of the nonlinear mixing between different modulated RF signals. As shown in Eq.(\ref{VS}) and (\ref{EST}), although the RF voltages are combined linearly, phase modulation is not a linear process. Therefore the quadratures of a certain channel will be effected by the nonlinear mixing from other channels. In addition, since the modulated information of each channel is independent, the extra source noise is also independent of the 
the Gaussian modulation of each channel. Hence it could be treated as a noise term in our analysis.

\begin{figure}
\centering\includegraphics[width=80mm]{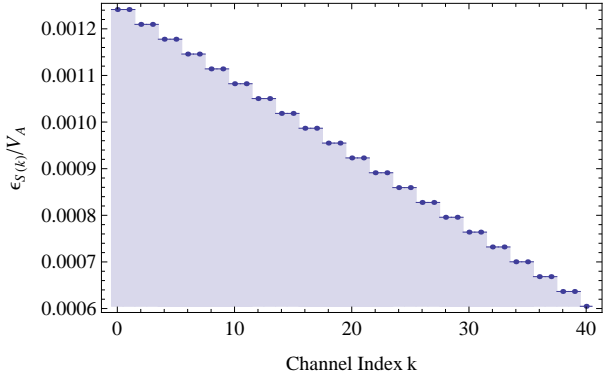}
\caption{(Color online) Source noise to modulation variance ratio $\epsilon_{S(k)}/V_A$ in terms of the channel index $k$. The total channel number is $N=40$ and the mean value of $m_k$ is $0.01$.}\label{MNK}
\end{figure}

Fig.\ref{MNK} shows the ratio  $\epsilon_{S(k)}/V_A$ in terms of the channel index $k$ when $N=40$. We find that the first channel ($k=1$) has the maximum value of $0.00124$ while the last channel ($k=N$) has the minimum one of $0.0006$. Other channels are placed between the first and last cases. When $N$ is large, e.g. $N=40$, the maximal value of $\epsilon_{S(k)}/V_A$ is approximately double of the minimal one.

Notice that although each quadrature of $V_A$ follows a Gaussian distribution, $\epsilon_{S(k)}$ is not a Gaussian noise. This situation is different from previous study about the source noise which assuming to be Gaussian\cite{YShen2010,YJShen2011}. According to Eq.(\ref{ESK}), since the extra source noise is proportional to $V_A$, it cannot be suppressed by increasing the variance of the modulation and then attenuating the state. So this noise must be taken into account in both theoretical analysis and actual experiments.

\begin{figure}
\centering
\includegraphics[width=80mm]{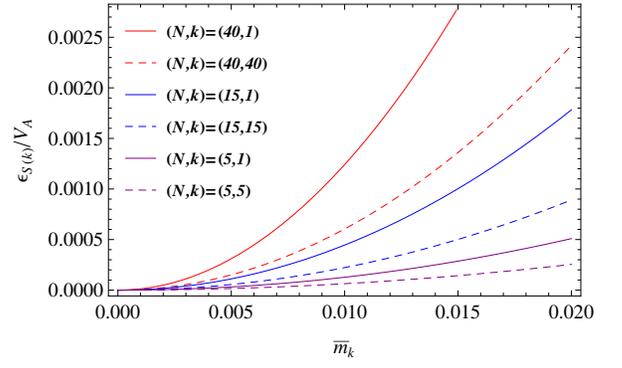}
\caption{(Color online) Source noise to modulation variance ratio $\epsilon_{S(k)}/V_A$ in terms of $\overline{m}_k$ of the first channel($k$=1) and the last channel($k$=$N$). Curves from top to bottom represent $(N,k)$=$(40,1)$, $(40,40)$, $(15,1)$, $(15,15)$, $(5,1)$ and $(5,5)$, respectively.}\label{ES}
\end{figure}

\section{Security of the multi-channel scheme}\label{S5}
\subsection{Entanglement-based scheme}

\begin{figure}
\centering
\subfigure[]{\includegraphics[width=89mm]{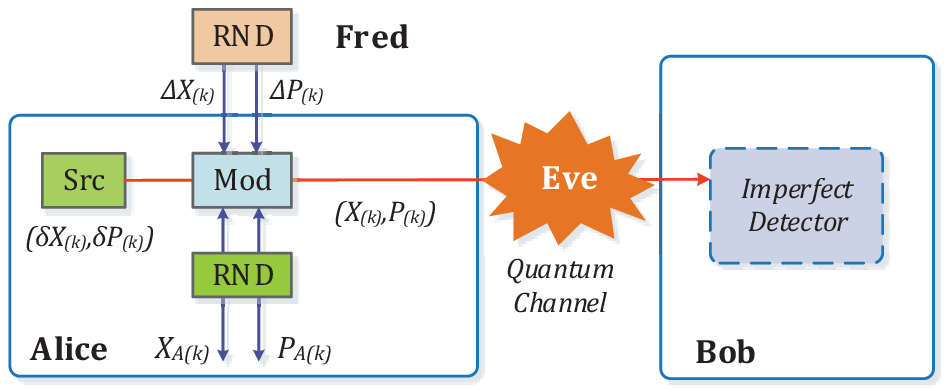}}
\subfigure[]{\includegraphics[width=92mm]{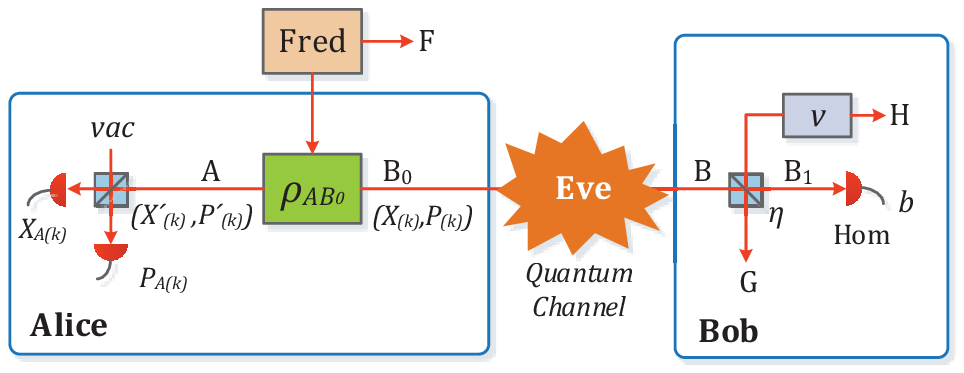}}
\caption{(Color online) The entanglement-based scheme of the $k$th channel. Fred is assumed to be a neutral party and Eve cannot benefit from neither Fred's information nor the imperfections of Bob's homodyne detector.}\label{PM-EB}
\end{figure}

In the prepare and measurement scheme of the $k$th channel:
\begin{eqnarray}
X_{(k)} &=& X_{A(k)}+\Delta X_{(k)}+\delta X_{(k)}\nonumber\\
P_{(k)} &=& P_{A(k)}+\Delta P_{(k)}+\delta P_{(k)},
\end{eqnarray}
where $\delta X_{(k)}$ and $\delta P_{(k)}$ are originated from shot noise and satisfy the relation that $\langle \delta X_{(k)}^2\rangle = \langle \delta P_{(k)}^2\rangle = 1$(in shot noise unit). $X_{A(k)}$ and $P_{A(k)}$ are Alice's modulated random numbers and satisfy the Gaussian distribution that $\mathcal{N}(0,V_A)$. $\Delta X_{(k)}$ and $\Delta P_{(k)}$ are the added non-Gaussian noise with variance $\epsilon_{S(k)}$. So the variance of $X_{(k)}$ and $P_{(k)}$ are:
\begin{eqnarray}
\langle  X_{(k)}^2\rangle = \langle P_{(k)}^2\rangle = V_A+1+\epsilon_{S(k)}.
\end{eqnarray}
And the conditional variances $V_{X_{(k)}|X_{A(k)}}$ and $V_{P_{(k)}|P_{A(k)}}$ are\cite{Grosshans2003}: 
\begin{eqnarray}
V_{X_{(k)}|X_{A(k)}} &=& \langle  X_{(k)}^2\rangle-\frac{\langle X_{(k)}X_{A(k)}\rangle^2}{\langle  X_{A(k)}^2\rangle} = 1+\epsilon_{S(k)}\nonumber\\
V_{P_{(k)}|P_{A(k)}} &=& \langle  P_{(k)}^2\rangle-\frac{\langle P_{(k)}P_{A(k)}\rangle^2}{\langle  P_{A(k)}^2\rangle} = 1+\epsilon_{S(k)}.
\end{eqnarray}

In the equivalent entanglement-based scheme, which is shown in Fig.\ref{PM-EB}(b), Fred generates a pure three-mode entanglement state $|\Psi_{ABF(k)}\rangle$ satisfied ${\rm tr}_F(|\Psi_{ABF(k)}\rangle\langle\Psi_{ABF(k)}|)=\rho_{AB_0(k)}$. The quadratures $(X_{(k)},P_{(k)})$ denote the state sent to Bob, and $(X'_{(k)},P'_{(k)})$ denote the state kept by Alice. Here we assume that  $(X_{(k)},P_{(k)})$ and $(X'_{(k)},P'_{(k)})$ satisfy the following relations:
\begin{eqnarray}\label{1}
\langle {X'}_{(k)}^{2}\rangle=\langle {P'}_{(k)}^2\rangle=V,\ \langle X_{(k)}^2\rangle=\langle P_{(k)}^2\rangle=V+\epsilon_{S(k)},
\end{eqnarray}
where $V=V_A+1$. According to the uncertainty relation\cite{Grosshans2003}, we have
\begin{equation}\label{limit}
|\langle X_{(k)}{X'}_{(k)} \rangle^2|\leq V(V+\epsilon_{S(k)})-\frac{V}{V+\epsilon_{S(k)}}.
\end{equation}
Since the three-party system ABF may not be maximally entangled, the correlation between modes $A$ and $B_0$ may not saturate the limit in Eq.(\ref{limit}). So it can be reasonably assumed that
\begin{equation}\label{2}
\langle X_{(k)}{X'}_{(k)}\rangle=\sqrt{V^2-1},\ \langle P_{(k)}{P'}_{(k)}\rangle=-\sqrt{V^2-1}.
\end{equation}
In the E-B scheme, when Alice takes a heterodyne detection on ${X'}_{(k)}$ and ${P'}_{(k)}$ simultaneously, the measurement values can be expressed as ${X'}_{A(k)}={X'}_{(k)}-\delta {X'}_{A(k)}$ and ${P'}_{A(k)}={P'}_{(k)}-\delta {P'}_{A(k)}$, where $\langle (\delta {X'}_{A(k)})^2\rangle=\langle (\delta {P'}_{A(k)})^2\rangle=1$. Alice's best estimation of $({X'}_{(k)},{P'}_{(k)})$ is denoted by $({X}_{A(k)},{P}_{A(k)})$ which satisfied\cite{Grosshans2003}
\begin{equation}
X_{A(k)}=\sqrt{\frac{V-1}{V+1}}{X'}_{A(k)},\ P_{A(k)}=-\sqrt{\frac{V-1}{V+1}}{P'}_{A(k)},
\end{equation}
and finally we have $\langle X_{A(k)}^2\rangle=\langle P_{A(k)}^2\rangle=V_A$ and $V_{X_{(k)}|X_{A(k)}}=V_{P_{(k)}|P_{A(k)}}=1+\epsilon_{S(k)}$, which has the identical results as in the P\&M scheme. If we hide the entanglement source and Alice's detection in a black box, the eavesdropper cannot distinguish which scheme is applied. So we can conclude that this E-B scheme is equivalent to the P\&M scheme.

In the P\&M scheme, Bob's imperfect detector is featured by a detection efficiency $\eta$ and the electric noise variance $v_{el}$. We could define an added noise $\chi_h$ referred to Bob's input as $\chi_h=(1-\eta+v_{el})/\eta$. In the E-B scheme, it is modeled by a beam splitter with transmission of $\eta$ coupled with an Einstein-Podolsky-Rosen(EPR) state of variance $v$. So in the E-B scheme, the added noise referred to Bob's input is $(1-\eta)v/\eta$. To make the detection-added noise equal, the variance $v$ should be chosen as $v = (1-\eta+v_{el})/(1-\eta)$.

\subsection{Security against collective attacks}

In this section, we consider the security of the multi-channel CV-QKD protocol with reverse reconciliation. According to the fact that coherent attacks are the most powerful eavesdropping attacks and are not more efficient than collective attacks\cite{Raul2006, Navascues2006}, we will analyze the security against collective attacks. 

Since the quadrature information of each channel is independent, we assume Eve's attacks of every channel also independent of others. Fred is supposed to be a neutral party which can not be controlled by the eavesdropper\cite{Huang2013}. This situation implies that both Alice and Eve cannot benefit from the information kept by Fred. Then the  secret key rate(bit/pulse) of the $k$th channel is expressed as
\begin{equation}\label{KK}
K_{(k)}=\beta I_{AB}-S_{BE},
\end{equation}
where $\beta$ is the efficiency of reverse reconciliation assumed to be constant for each channel. $I_{AB}$, which represents the Shannon mutual information between Alice and Bob, can be derived from Bob's measured variance $V_{B_1}$ and the conditional variance $V_{B_1|A}$ as
\begin{equation}
I_{AB}=\frac{1}{2}\log_2\frac{V_{B_1(k)}}{V_{B_1|A}}=\frac{1}{2}\log_2\frac{V+\epsilon_{S(k)}+\chi_{tot}}{1+\epsilon_{S(k)}+\chi_{tot}},
\end{equation}
where $\chi_{tot} = \chi_{line}+\chi_{h}/T$ and $\chi_{line}=1/T-1$. $T$ is the transmission efficiency of the quantum channel and can be evaluated by the transmission distance $L$ by  $T = 10^{-0.02L}$. 
Eve's information on Bob's measurement is given by the Holevo bound\cite{Holevo1973}:
\begin{equation}
S_{BE}=S(\rho_E)-S(\rho_{E|B=b}),
\end{equation}
where $S(\cdot)$ denotes the von-Neumann entropy and $b$ represents the measurement result of Bob. Here we consider Alice and Fred together as a larger state $A'$. Since the fact that Eve has the ability to purify the system $A'B$, we have $S(\rho_{E})=S(\rho_{A'B})$. After Bob's measurement, the global pure state collapses to $\rho_{A'EGH|b}$, so  $S(\rho_{E|b})=S(\rho_{A'GH|b})$. Notice that the state $\rho_{A'GH|b}$ is determined by state $\rho_{A'B}$\cite{YShen2010}. According to the optimality of Gaussian attacks\cite{Navascues2006,Raul2006}, $S_{BE}$ reaches its maximum when the state $\rho_{A'B}$(i.e. $\rho_{FAB}$) is Gaussian. Then the Eve's information can be bounded by
\begin{equation}
S_{BE}\leq S_{BE}^G=S(\rho_{FAB}^G)-S({\rho^{G}_{FAGH|b}}),
\end{equation}
where $\rho_{FAB}^G$ is a Gaussian state with the covariance matrix\cite{Huang2013}
\begin{equation}
\gamma_{FAB}^G=\left[
\begin{array}{ccc}
F_{11} & F_{12} & F_{13} \\
F_{21} & V\mathbb{I}_2 & \sqrt{T(V^2-1)}\sigma_z\\
F_{31} & \sqrt{T(V^2-1)}\sigma_z & T(V+\epsilon_{S(k)}+\chi_{line})\mathbb{I}_2\\
\end{array}
\right],
\end{equation}
which is identical to the covariance matrix of $\rho_{FAB}$. $\mathbb{I}_{2} = \left [
\begin{smallmatrix}
1&0\\
0&1\\
\end{smallmatrix}
\right]$ and $\sigma_{z} = \left [
\begin{smallmatrix}
1&0\\
0&-1\\
\end{smallmatrix}
\right]$.
$F_{m,n}$ represents the unknown $2\times2$ matrix describing either $F$ or its correlations with $AB$. Although the entropy of $\rho_{FAB}^G$ can not be calculated directly, there exists another Gaussian state $\rho'^G_{FAB}$ with the covariance matrix $\gamma'^G_{FAB}$
\begin{equation}
\gamma'^G_{FAB}=\left[
\begin{array}{ccc}
\mathbb{I}_2 & 0 & 0 \\
0 & V'\mathbb{I}_2 & \sqrt{T(V'^2-1)}\sigma_z\\
0 &  \sqrt{T(V'^2-1)}\sigma_z & T(V'+\chi_{line})\mathbb{I}_2\\
\end{array}
\right],
\end{equation}
where $V'=V+\epsilon_{S(k)}$. 
The reduced state $\rho'^G_{B}=\text{tr}_{FA}(\rho'^G_{FAB})$ is identical to the reduced state $\rho^G_{B}=\text{tr}_{FA}(\rho^G_{FAB})$, therefore $\rho^{G}_{{FAB}}$ can be changed to  $\rho'^G_{FAB}$ through a unitary transformation $U_{FA}$\cite{Chuang2000}. Then we have  $S(\rho'^G_{FAB})=S(\rho^G_{FAB})$. Similarly, the conditional state $\rho^{G}_{FAGH|b}$ can be transforms into the  $\rho'^{G}_{FAGH|b}$ through $U_{FA}$, then $S(\rho'^{G}_{FAGH|b})=S(\rho^{G}_{FAGH|b})$. Therefore we have
\begin{equation}
S_{BE} \leq S^G_{BE}=S(\rho'^G_{FAB})-S({\rho'^{G}_{FAGH|b}}),
\end{equation}
and the lower bound $\widetilde{K}_{(k)}$ of the secret key rate  can be expressed as
\begin{eqnarray}\label{boundk}
\widetilde{K}_{(k)}&=&\beta I_{AB}-S^G_{BE},\nonumber\\
&=&\beta I_{AB}-\sum_{j=1}^3G\left(\frac{\lambda_j-1}{2}\right)+\sum_{j=4}^7G\left(\frac{\lambda_j-1}{2}\right),
\end{eqnarray}
where $G(x)=(x+1)\log_2(x+1)-x\log_2 x$. The first three sympletic eigenvalues $\lambda_{1,2,3}\geq 1$, which are derived from the covariance matrix $\gamma'^G_{FAB}$, can be expressed as
\begin{equation}\label{lam123}
\lambda_{1,2}^2=\frac{1}{2}(A\pm\sqrt{A^2-4B}),\ \lambda_3=1,
\end{equation}
where 
\begin{eqnarray}\label{AB}
A &=& (V+\epsilon_{S(k)})^2-2T[(V+\epsilon_{S(k)})^2-1]\nonumber\\
&&+T^2(V+\epsilon_{S(k)}+\chi_{line})^2\nonumber\\
B &=& T^2[1+(V+\epsilon_{S(k)})\chi_{line}]^2.
\end{eqnarray}
The symplectic eigenvalues $\lambda_{4,5,6,7}\geq 1$ can be obtained from  the covariance matrix
 $\gamma'^{G}_{FAGH|b}$ and have the following form
\begin{eqnarray}\label{lam4567}
\lambda_{4,5}^2=\frac{1}{2}(C\pm\sqrt{C^2-4D}),\ \lambda_{6,7}=1,
\end{eqnarray}
where 
\begin{eqnarray}\label{CD}
C &=& \frac{A\chi_{h}+(V+\epsilon_{S(k)})\sqrt{B}+T(V+\epsilon_{S(k)}+\chi_{line})}{T(V+\epsilon_{S(k)}+\chi_{line})+\chi_{h}}\nonumber\\
D &=& \frac{\sqrt{B}(V+\epsilon_{S(k)})+B\chi_{h}}{T(V+\epsilon_{S(k)}+\chi_{line})+\chi_{h}},
\end{eqnarray}
where $A$ and $B$ are given in Eq.(\ref{AB}). Based on Eq.(\ref{lam123}), (\ref{AB}), (\ref{lam4567}) and (\ref{CD}), we can calculate the asymptotic lower bound of the secret key rate in Eq.(\ref{boundk}) against collective attacks.

\section{Simulation and discussion}\label{S6}
\begin{figure}
\includegraphics[width=89mm,height=60mm]{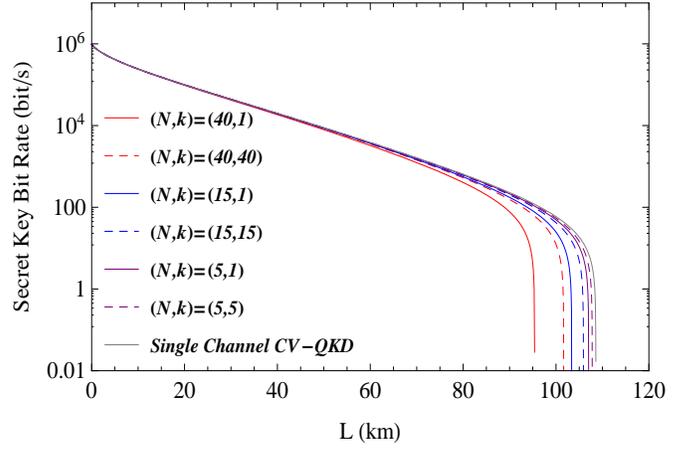}
\caption{(Color online) The secret key bit rate $R_{(k)}$ as a function of transmission distance $L$ of the first and last channel. Curves from left to right are $(N,k)=(40,1)$, $(40,40)$, $(15,1)$, $(15,15)$, $(5,1)$, $(5,5)$ and single channel case.}\label{KEY-L}
\end{figure}
\begin{figure}
\includegraphics[width=89mm,height=60mm]{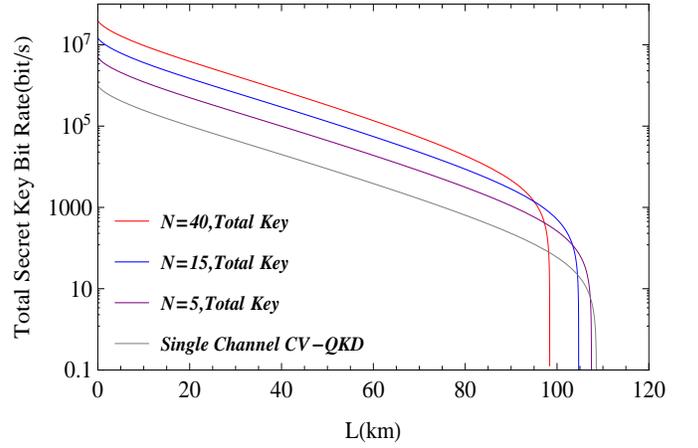}
\caption{(Color online) The total secret key bit rate $R_{tot}$ as a function of transmission distance $L$  in cases of different channel plans. From top to bottom: $N=40$, $15$, $5$ and signal channel case. }\label{TOTAL-KEY}
\end{figure}
For a given total channel number $N$, the bit rate of the secret key of the $k$th channel is given as
\begin{equation}
R_{(k)}=f_{\text{rep}}\widetilde{K}_{(k)},
\end{equation}
where $\widetilde{K}_{(k)}$ is the secret key bit per pulse and can be evaluated using Eq.(\ref{KK}). 
We assume the system repetition rate $f_{\text{rep}}$, quantum efficiency $\eta$ and the electronic noise of homodyne detector $v_{el}$ as $f_{\text{rep}}=1$MHz, $\eta=0.552$ and $v_{el}=0.015$, corresponding to the typical experimental parameters\cite{Jouguet2013}. The excess noise $\epsilon$ is assumed to be $\epsilon=0.02$ which is a conservative value\cite{Lodewyck2007}. The reverse reconciliation is set to $\beta=0.93$ which is an achievable value with existing techniques\cite{Jouguet2011}. We choose $V_A=10$ as the modulation variance at Alice's side.
\begin{figure}
\subfigure[]{\includegraphics[width=89mm,height=60mm]{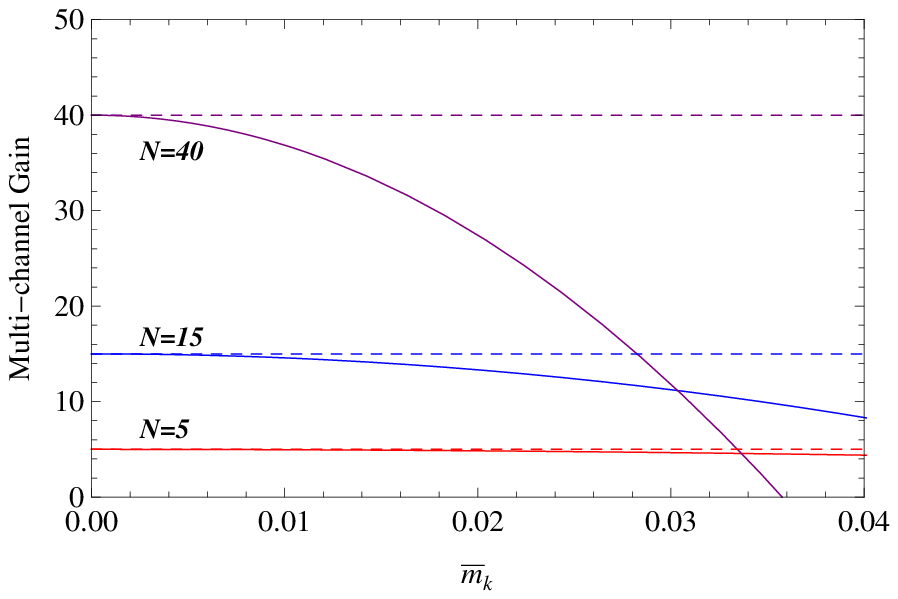}}
\subfigure[]{\includegraphics[width=89mm,height=60mm]{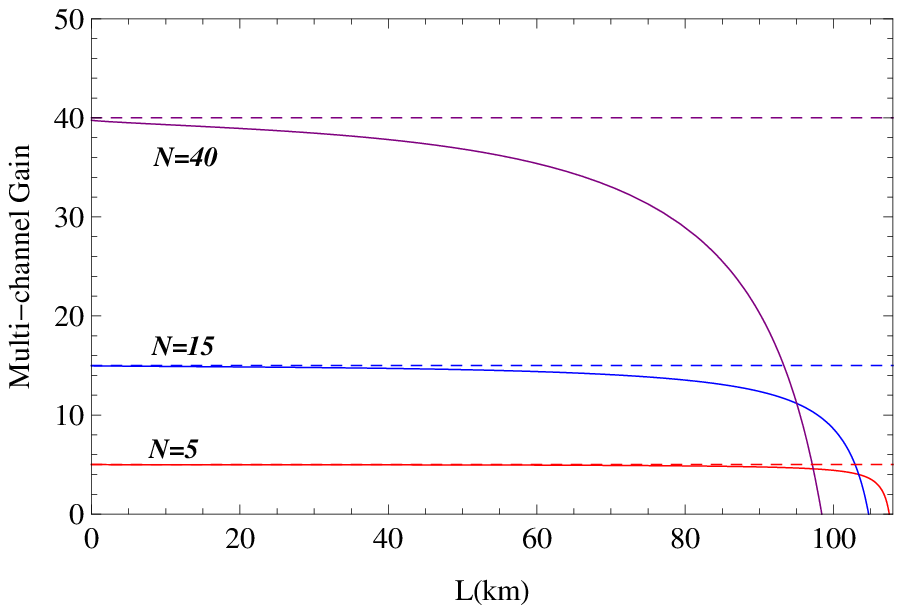}}
\caption{(Color online)The multi-channel gain of the system. (a) is the function of $\overline{m}_k$ at the distance of $L=50$km. (b) is the function of transmission distance when $\overline{m}_k=0.01$. Curves from top to bottom are $N=40$, $15$ and $5$. Dashed lines act as a reference. Other parameters: $\beta=0.93$, $\epsilon=0.02$, $\eta=0.552$, $v_{el}=0.015$ and $V_A=10$.}\label{MCGAIN}
\end{figure}

According the Fig.\ref{MNK}, since the first and last channels have the maximal and minimal extra source noise, we only need to evaluate the secret key of these two channels while the other channels are placed between them. Fig.\ref{KEY-L} shows the secret key bit rate for the first($k=1$) and last($k=N$) channel in cases of $N=5,15,40$. For each case, the $k=N$ channel always performs best among all the channels in both secret key rate and the maximum transmission distance while the $k=1$ channel performs the worst. This is because the extra source noise introduced by intermodulation is a decreasing function of the channel index $k$ as shown in Fig.\ref{MNK} and a increasing function of the total channel number $N$ as shown in Fig\ref{ES}. Due to the extra source noise, the maximum transmission distance and the secret key rate of the subcarrier channels are smaller than single-channel CV-QKD.

The total secret key bit rate can be defined as a sum of key rate of each channel
\begin{equation}
R_{tot}=\sum_{k=1}^{N}R_{(k)}.
\end{equation}
$R_{tot}$ as a function of transmission distance $L$ is demonstrated in Fig.\ref{TOTAL-KEY}. The total secret key bit rate is considerably increased with the channel numbers $N$ from zero to 80km. Interestingly, the maximum transmission distance is decreased $N$. We also find that the high-count channel system(e.g.$N=40$) may performs worse than the mid-count and low-count channel systems in certain distance ranges(90$\sim$110km). This is mainly due to the fact that $R_{tot}$ of the  high-count system starts to go down and falls to zero rapidly, while at this distance the low-count system still has positive key rate. 

In order to estimate the increment on the secret key bit rate of multi-channel protocol, we define the multi-channel gain as
\begin{equation}
G_{M}=\frac{R_{tot}}{R_{sc}}=\frac{1}{R_{sc}}\sum_{k=1}^{N}R_{(k)},
\end{equation}
where $R_{sc}$ represents the secret key bit rate of a single-channel CV-QKD with the identical parameters as those in the multi-channel scheme. Fig\ref{MCGAIN}(a) shows the evolution of $G_M$ as a function of the mean value of $m_k$ at the distance of $L=50$km. For $\overline{m}_k\leq0.005$, the effect of nonlinear signal mixing can be neglected and the multi-channel gain is almost identical to the number of channels($N$). With the increase of $\overline{m}_k$, $G_{M}$ is reduced, so $G_M\leq N$. Fig.\ref{MCGAIN}(b) demonstrates the relation between $G_M$ and the transmission distance $L$ with $\overline{m}_k=0.01$. $G_M$ descends rapidly with the increment of $L$ and finally falls to zero when $R_{tot}$ reaches its maximum transmission distance. In short distance ranges(e.g. $L\leq 30km$), the multi-channel system has a great improvement on the total secret key bit rate with a gain of about $N$ times.

\section{Conclusions}\label{S7}

In summary, we present a scheme for continuous variable quantum key distribution using the subcarrier multiplexing technique in microwave photonics. We study the generation of both the subcarrier signal and the local oscillator light. We also analyze the influence of nonlinear signal mixing and the extra source noise due to intermodulation. We find the extra source noise is non-Gaussian distributed and proportional to the modulation variance $V_A$. Then we investigate the security against collective attacks and evaluate the lower bound for the secret key rate.  Our results show that by using this multiplexing technique, the maximum transmission distance of each channel will decreased slightly, while the total secret key rate could have a considerable improvement.

We also notice that our scheme could be used for key distribution for multi users. As shown in Sec.\ref{S4},  each channel generates an independent secret key at the same time, meaning the key distribution in the multi-channel system is parallel. This scheme could be used for one Alice to distribute keys to several Bobs. Limited by the electro-optic modulators, the bandwidth occupied by the multi-channel scheme is under 100GHz, which is smaller than the interval of WDM devices. So several multi-channel CV-QKD systems with different central frequencies could be combined with WDM devices, resulting in a potential to construct a CV-QKD network.

Future work will be the experiments of generating the multi-channel signals and the demonstration of the multi-channel CV-QKD system. 
This work is supported by the National Natural Science Foundation of China (Grant Nos. 61170228), and China Postdoctoral Science Foundation (Grant No. 2013M540365).

\end{document}